# Clay-Based Hybrid Membranes for Antibacterial and Antifungal Wound Healing Applications

*Suvash Ghimire[1], Yi Wu[2], Manjyot Kaur Chug[2], Elizabeth J. Brisbois[2], Kyungtae Kim[3], Kausik Mukhopadhyay[1*]*

[1]Materials Science and Engineering, University of Central Florida, Orlando, FL, 32816, United States

[2]School of Chemical, Materials, and Biomedical Engineering, University of Georgia, Athens, GA, 30602, United States

[3]Materials Physics and Applications Division, Center for Integrated Nanotechnologies, Los Alamos National Laboratory, Los Alamos, NM, 87545, United States

**Abstract**:

Microbes and pathogens play a detrimental role in healing wounds, causing infections like impetigo through bodily fluids and skin and entering the bloodstream through the wounds, thereby hindering the healing process and tissue regeneration. Clay, known for its long history of therapeutic use, has emerged as one of the most promising candidates for biomedical applications due to its non-toxic nature, high surface area, ubiquity, and excellent cation exchange capacity. This study demonstrates an innovative approach to engineering an organo-functionalized, infection-resistant, easy-to-use bandage material from clay, an environmentally benign and sustainable material. The hybrid membranes have been developed using clays, zwitterions, silver ions, and terbinafine hydrochloride (TBH) to impart antibacterial and antifungal efficacy. A critical aspect of this study is embedding organic molecules and metal ions with the clays and releasing them to resist the growth and kill the pathogens. The antimicrobial efficacy of the membranes has been tested using a zone of inhibition study against the most common microbes in skin wounds, *viz. S. aureus*, *E. coli*, and *C. albicans*. Our study demonstrates the potential of these hybrid clay membranes as a cost-effective, scalable, and effective solution for treating microbial infections and instills newer avenues for point-of-care wound-healing treatments.

## 1. Introduction

Microbial infections from wounds, burns, and foodborne illnesses are a compounding and challenging issue that accounts for 12% of fatalities globally, of which the highest mortality rates are impacted in low- and middle-income countries[1]. In one of the most extensive studies carried out and published recently, it has been reported that five pathogens – *Staphylococcus aureus*, *Escherichia coli*, *Streptococcus pneumoniae*, *Klebsiella pneumoniae*, and *Pseudomonas aeruginosa* accounted for almost 55% of the 7.7 million deaths, with *S. aureus* associated with over 1.1 million deaths[2]. Moreover, infectious syndromes can be highly contagious among humans and animals, often forcing the poultry and farming industry to sacrifice millions of poultry suffering from bacterial and viral infections[3].

Superficial wounds harm the upper layers of the skin, while acute wounds are more severe and can lead to bacterial and fungus infections[4]. The most common bacteria in the human body are *Staphylococcus aureus* (*S. aureus*) and *Escherichia coli* (*E. coli*)[5]. For instance, *S. aureus* is the most common source of infection in surgical disease[6]. These bacteria reside harmlessly in the human body; however, when entering the bloodstream through a wound, they can cause significant infections and release enzymes that aid in transmitting infections to adjacent tissues[7].

Although bacteria are typically considered the primary contributors to chronic nonhealing wounds, fungi also play a significant role in these cases[8]. Fungal infections pose the risk of mortality and morbidity in critical care settings, especially in chronic and surgical wounds[6]. It is reported that almost a quarter of chronic wounds may be infected with one or more fungi[9]. The most common fungal pathogen in chronic wounds is yeast-like *Candida* species, particularly Candida albicans (*C. albicans)*[10]. *C. albicans* is a fungal species that naturally occurs in the human body, particularly in the gastrointestinal tract, reproductive tract, oral cavity, and skin[11]. In skin wounds, fungal infections complicate the condition by promoting the formation of biofilms that are resistant to conventional treatments[12].

Current wound dressing materials available in the market include sponges, bioactive glasses, carbon-based supports, hydrogels, hydrocolloids, films, and fibers made from synthetic and natural materials[13–15]. In many systems, antimicrobial agents, such as metal nanoparticles, antibiotics, etc., have been incorporated into wound dressings to enhance antibacterial efficacy[14–16]. Although several synthetic materials have been developed to combat pathogens, supplementary and cost-effective strategies are sought after in wound care and treatment these days, which requires developing alternative wound healing solutions from environmentally friendly and sustainable materials that can address toxicity, ease of use in inclement conditions, packaging, and poor biodegradability[17].

Clays have been used for therapeutic and biomedical applications due to their excellent biocompatibility, natural abundance, surface area, efficient oxygen and moisture transport through their pores, and presence of elements such as iron, magnesium, and potassium that aid healing.[20,21] Due to their unique structure and chemical properties, clays, such as bentonite and sepiolite, can be easily functionalized with organic molecules and polymers for various applications, including antimicrobial applications[18,19].

Antibiotics are the most often utilized materials for antibacterial wound dressing applications, followed by metal ions and metal oxide nanoparticles[16]. While silver salts and nanoparticles have been the most common for antibacterial treatments[20], azoles, polyenes, echinocandins, and allylamines are prevalent in treating fungal infections, such as dermatophyte[10]. One such example, Terbinafine hydrochloride (TBH), is an orally and topically active allylamine known for its activity against various fungi, even at low doses[21]. Its capability to fight against many dermal fungi, such as dermatophytes, dimorphic fungi, dematiaceous fungi, and yeast, makes it a suitable drug for transdermal fungal treatment[21]. With the correct dosage over an appropriate period, silver is an antiseptic that can provide efficient cidal activity and efficacy against various wound pathogens. Besides, there is strong evidence that nanocrystalline silver and silver ions can minimize transmission of antibiotic-resistant organisms through proper use and stewardship of local wound infections to reduce the need for antibiotic therapy, which helps reduce the time and cost of treatment[22].

Herein, we aim to showcase a simple solution for an affordable antimicrobial treatment by engineering unique hybrid clay composite membranes using environmentally benign materials, such as clays, betaine, silver ions, and TBH, with excellent antibacterial and antifungal efficacy. Two varieties of clay, bentonite, and sepiolite were used as host materials to synthesize the composite samples. Betaine, a non-toxic zwitterion, acts as a tethering agent and facilitates the adsorption of antimicrobial agents to the clay's interlayer or surface. Free-standing Ag-betaine-bentonite hybrid clay composite membranes were synthesized by infusing silver ions into bentonite clay. These composites were effective against gram-positive and gram-negative bacteria (*S. aureus* and *E. coli*) but not against fungi. TBH-betaine-bentonite clay composites were synthesized for antifungal application. Subsequently, Ag-TBH-betaine-bentonite clay composites were synthesized by infusing silver ions and TBH into the bentonite clay to achieve activity against bacteria and fungi. To validate the importance of the betaine-functionalization of clay in synthesizing the composites, we developed another system, an Ag-betaine-sepiolite clay composite, for bacterial treatment and an Ag-TBH-betaine-sepiolite clay composite for bacterial and fungal treatment. The choice for betaine as the zwitterion in this study stems from our recently published work, where we explored the effect of functionalizing bentonite clay with betaines of variable carbon chain lengths to study the rheological properties of clay slurries to analyze their interactions in suspension[23]. The results showed that these zwitterion-functionalized clays exhibit higher viscosity, storage moduli, and flow stresses due to the formation of three-dimensional networks and increased aggregation caused by intercalation[23]. Using rheological studies, we investigated the interaction between the clay and zwitterions in slurries and how the interactions impact the structure and properties of bentonite clay during functionalization with betaines. We concluded that tweaking the rheological properties of clays by intercalating small, pH-sensitive, environment-friendly betaine molecules such as trimethyl glycine offers significant improvements rheologically compared to pristine, unmodified clays and is a promising alternative for non-toxic rheological additive, which contrasts with long-chain surfactants currently being used to make most functionalized clays.

## 2. Experimental Section

### 2.1. Materials and Supplies

Bentonite clay, sepiolite clay, and phosphate buffer saline (PBS) were purchased from Sigma Aldrich (St. Louis, MO). Betaine, silver nitrate, and terbinafine hydrochloride (TBH) were purchased from Fisher Scientific (Hampton, NH) and used for the synthesis without further purification. Deionized water was used for all experiments. The bacterial strains *S. aureus* (ATCC 6538) and *E. coli* (ATCC 25922), as well as the fungal strain *C. albicans* (ATCC MYA 4441), were purchased from the American Type Culture Collection (ATCC). All the buffers and media used in microbial culture were sterilized in an autoclave at 121 °C and 15 psi for 30 minutes before using for experiments.

### 2.2. Synthesis of Clay Composite

### 2.2.1. Bentonite Clay Composite

**Betaine-Bentonite Clay Composite (Membrane):**

The betaine-bentonite clay composites were synthesized by intercalating a zwitterionic surfactant, betaine, into the clay galleries. In a typical synthesis, 3.3 g of Na-bentonite clay was dispersed in 50 ml of deionized (DI) water. The mixture was shaken in a mechanical shaker for 24 hours to allow the silicate layer to swell. The clay slurry was centrifuged thrice at 8000 rpm and washed with DI water to remove impurities. After centrifugation, the clay residue was mixed with 0.58 g of betaine and 10 ml of DI water and mechanically mixed for another 24 hours. Finally, the prepared clay slurry was coated over a glass plate using an automatic film coater at 2mm/s. The coated glass plates were kept in the oven for drying at 60 ºC for 4 hours. After drying, the resulting clay composite membrane was peeled off the glass plate and named Betaine/Bentonite Clay Composite Membrane BB-CC (M), which served as the control sample. The samples were cut into 6 mm discs for antibacterial testing. The weight and thickness of the composite membranes were 2.5 mg and 50 μm, respectively.

**Ag-Betaine-Bentonite Clay Composite (Membrane):**

To begin with, the Ag-Betaine-Bentonite Clay Composites (Ag-BB-CC) membranes were synthesized using an ion exchange mechanism, where $Na^+$ ions in Na-bentonite clay were cation-exchanged with $Ag^+$ ions. Four Ag-BB-CC samples were prepared with varying amounts of silver nitrate ($Ag^+$ conc.) to investigate the effect of $Ag^+$ ion concentration on the antibacterial activity of the Ag-BB-CC samples. Typically, 3.3 g of bentonite clay and 0.58 g of betaine were used in each sample, and varied amounts of silver nitrate (0.3 g, 0.6 g, 0.9 g, and 1.2 g) were used.

The synthesis involved adding pre-weighed amounts of silver nitrate in bentonite clay in DI water and shaking the slurries for 24 hours to facilitate the ion exchange. This was followed by centrifuging the mixtures at 8000 rpm and washing them with copious amounts of deionized water to remove impurities and excess or unexchanged ions ($Na^+$ and $Ag^+$). The Ag-clay slurry residue was then transferred to a new container and mixed with 0.58 g of betaine and 10 ml of DI water, followed by shaking for another 24 hours. The resulting slurry was carefully poured onto a glass plate, and the rate and thickness were controlled using an automatic film coater. It was dried at 60°C for 4 hours, then peeled off the silver-exchanged clay membrane from the glass plate, resulting in the Ag-Bentonite-Betaine clay composite membrane (Ag-BB-CC (M)). The composite membranes were synthesized with varying concentrations of silver nitrate in the slurries that were designated as Ag (1)-BB-CC (M), Ag (2)-BB-CC (M), Ag (3)-BB-CC (M), and Ag (4)-BB-CC (M). The weight and thickness of the composite membranes were approximately 2.5 mg and 50 μm, respectively.

**TBH-Betaine-Bentonite Clay Composite (Pellet):**

TBH-Betaine-Bentonite Clay composites were synthesized by adding 3.3 g of bentonite clay in 50 ml of DI water. The mixtures were agitated in a mechanical shaker for 24 hours, followed by centrifugation at 8000 rpm three times and washing with copious amounts of DI water. The samples were transferred to a separate container and mixed with pre-made TBH solutions in methanol. The clay-TBH slurries were then allowed to shake for another 24 hours at room temperature, coated on glass plates, and dried in a vacuum oven at 60 °C for 24 hours. The powder samples were then collected and compacted into pellets using a pellet maker. The pellet samples were named TBH (1)-BB-CC (P), TBH (2)-BB-CC (P), and TBH (3)-BB-CC (P). The concentrations of TBH in TBH (1)-BB-CC, TBH (2)-BB-CC, and TBH (3)-BB-CC were 50 mg, 100 mg, and 150 mg in 5 ml of methanol, respectively. The weight and thickness of the composite pellets were 200 mg and 2.6 mm, respectively.

**Ag-TBH-Betaine-Bentonite Clay Composite (Pellet):**

Ag-TBH-Betaine-Bentonite Clay composites were synthesized by mixing 0.3 g of silver nitrate and 3.3 g of bentonite clay in 50 ml of deionized water. A TBH solution was prepared by dissolving 50 mg of TBH in 5 ml of methanol, which was then added to the Ag-clay mixture. This resulting mixture was mechanically agitated for 24 hours in a shaker and then centrifuged. Subsequently, 0.58 g of betaine was added, and the agitated slurry was then coated on a glass plate and dried in an oven at 60°C for 4 hours. The dried powder samples were collected and compacted into 6 mm pellets for antimicrobial testing. The weight and thickness of the composite pellets were 200 mg and 2.6 mm, respectively, and were labeled Ag-TBH-BS-CC (P).

### 2.2.2. Sepiolite Clay Composite

**Betaine-Sepiolite Clay Composite (Membrane):**

Betaine-sepiolite composite membranes were synthesized using sepiolite clay and betaine using the procedure mentioned in section 2.1.1.1. The composite is named BS-CC (M). The composite membrane was used as a control sample for the sepiolite clay composite. The weight and thickness of the composite membranes were 5 mg and 70 μm, respectively.

**Ag-Betaine-Sepiolite Clay Composite (Membrane)**

Ag-Betaine-Sepiolite Clay Composite Membranes were prepared using 0.3 g of silver nitrate, 3.3 g of sepiolite clay, and 2.5 g of betaine, following the procedure described in section 2.2.1.2.

**Ag-TBH-Betaine-Sepiolite Clay Composite (Pellet)**

Ag-TBH-Betaine-Sepiolite Clay composite samples were synthesized using 0.3 g of silver nitrate, 3.3 g of sepiolite clay, and 2.5 g of betaine, following the procedure outlined in section 2.2.1.4. The resulting powder samples were collected and compacted into pellets with a diameter of 6 mm for antimicrobial testing. The composite pellets weighed 200 mg, had a thickness of 2.6 mm, and were designated as Ag-TBH-BS-CC (M). Details of the samples are provided in Table 1.

**Table 1. Clay-composite sample Information.**

| Sample | Abbreviation | Physical form | Application |
|---|---|---|---|
| Betaine-Bentonite Clay Composite | BB-CC (M) | Membrane | Control |
| Ag-Betaine-Bentonite Clay Composite | Ag (1)-BB-CC (M) | Membrane | Antibacterial |
| | Ag (2)-BB-CC (M) | Membrane | Antibacterial |
| | Ag (3)-BB-CC (M) | Membrane | Antibacterial |
| | Ag (4)-BB-CC (M) | Membrane | Antibacterial |
| TBH-Betaine-Bentonite Clay Composite | TBH (1)-BB-CC (P) | Pellet | Antifungal |
| | TBH (2)-BB-CC (P) | Pellet | Antifungal |
| | TBH (3)-BB-CC (P) | Pellet | Antifungal |
| Ag-TBH-Betaine-Bentonite Clay Composite | Ag-TBH-BB-CC (P) | Pellet | Antibacterial, Antifungal |
| Betaine-Sepiolite Clay Composite | BS-CC (M) | Membrane | Control |
| Ag-Betaine-Sepiolite Clay Composite | Ag-BS-CC (M) | Membrane | Antibacterial |
| Ag-TBH-Betaine-Sepiolite Clay Composite | Ag -TBH -BS-CC (P) | Pellet | Antibacterial, Antifungal |

### 2.3. Testing and Characterization

#### 2.3.1. Powder X-ray diffraction (XRD) and small-angle X-ray scattering (SAXS)

Powder X-ray diffraction (XRD) of clay composites was performed on a PANalytical Empyrean X-ray diffractometer equipped with a copper X-ray tube ($\lambda_{Cu\ K\alpha}$ = 0.1544426 nm) and X'Celerator multi-element detector for rapid data acquisition. The voltage and current of the X-ray tubes were 45 kV and 40 mA, respectively. Diffraction data were collected from 2° to 42° (2$\theta$) with a step size increment of 0.08°. XRD was used to measure the interlayer spacing of the clay, and the interlayer spacing of both unmodified and modified clay was determined by measuring the $d_{001}$ reflection using Bragg's law, i.e., $2d\ sin\theta = n\lambda$.

Small-angle X-ray scattering (SAXS) was performed using a Xenocs Xeuss 3.0 small-angle/wide-angle X-ray scattering instrument equipped with a Cu X-ray source and an Eiger2 R 1M Dectris detector. The sample-to-detector distance was varied from 7.2 mm to 1650 mm. Upon measurements, two-dimensional (2D) scattering patterns were obtained, and azimuthal integration of these scattering patterns resulted in one-dimensional (1D) scattering profiles, $I(q)$, versus scattering vector $q = 4\pi\lambda^{-1} sin\theta$. The nanostructures were determined using relative Bragg peak positions of the scattering profiles[24].

#### 2.3.2. ATR-IR

Infrared (IR) spectra were recorded using a Cary 60 Model ATR-IR spectrometer. Spectra were collected in the 4000-500 cm-1 range, with 128 scans for each specimen at a resolution of 4 cm$^{-1}$.

#### 2.3.3. X-ray Photoelectron Spectroscopy (XPS)

The chemical composition of the prepared composites was investigated using an X-ray photoelectron spectrometer (XPS, Physical Electronics 5400 ESCA) with an Al-monochromatic

X-ray source (20 mA, 15 kV). Thermo Avantage Peakfit® software was used to obtain the deconvoluted peaks. After background subtraction, the spectra were fitted with a Gaussian-Lorentzian function.

### 2.3.4. Thermogravimetric Analysis (TGA)

Thermogravimetric analysis of organofunctionalized clay composites was conducted using PerkinElmer STA 6000 equipment. In a typical experiment, the samples were placed in a ceramic pan and heated at 20 °C/min in an $N_2$ atmosphere (20 ml/min). The tests were carried out at temperatures ranging from 20 to 420 °C.

### 2.3.5. TBH Leaching

TBH-containing samples (TBH-BB-CC (P) and Ag-TBH-BS-CC (P)) were dialyzed against D.I. water for 48 h using 3500 Dalton MWCO dialysis membrane tubing. Dialysis solution was obtained at 24 and 48 h, and the absorption was read at $\lambda_{max}$ = 283 nm using a UV-Vis spectrophotometer (Thermo-scientific Genesys 10S UV-Vis). The calibration curve of TBH was prepared using known TBH concentration dissolved in 1 mL of PBS to interpolate the absorbance measurement recorded from the UV-Vis converted to TBH concentration in the solution. TBH calibration is shown in supplementary Figure S10.

### 2.3.6. Inductively Coupled Plasma-Optical Emission Spectroscopy (ICP-OES)

The ICP-OES analyses were carried out using vaporization on PerkinElmer Avio 500 instruments. Optima 3300DV, 4300DV, 5300, and 8300 were used for emission wavelength measurement. Before inspection, the samples were fused with sodium peroxide over a Bunsen burner. After being dissolved in water, the fused samples were acidified with nitric and hydrochloric acid. The prepared samples were nebulized, and the resulting aerosol was transferred to the plasma torch, which used radio frequency inductive coupling plasma to generate the element's spectra. The spectra were dispersed using a grating spectrometer, and the intensity of emission lines was monitored using a photosensitive detector. Each solution had a final volume of 100 mL. The plasma, auxiliary, and nebulizer gas rates were 10/min, 0.2 L/min, and 0.7 L/min, respectively.

For ICP-MS (Mass Spectroscopy) analyses, the samples were incubated in PBS for 24 h, and the leachate was obtained for ICP-MS measurement. Nitric acid oxidizes organic matter to $CO_2$ and NO while forming soluble nitrates with most elements before being analyzed. Concentrations reported in mg/kg refer to milligrams of metal per kilogram of Ag-betaine-bentonite clay samples sample (ppm).

### 2.3.7. Antimicrobial Efficacy of Clay Composites

The antibacterial efficacy of the clay samples was evaluated against *S. aureus*, *E. coli*, and *C. albicans* using a zone of inhibition (ZOI) study. For this, the colonies of each strain were inoculated into either LB media or YEM media at 150 rpm at 37 °C. The optical density (O.D.) of the culture was recorded using a UV-vis spectrophotometer (Cary-60, Agilent Technologies) at 600 nm ($OD_{600}$) wavelength. All samples were sterilized for 20 mins under UV light before bacteria exposure ($n \geq$ three each). The final O.D. of the culture was adjusted to 0.1, and 50 µL of bacteria culture was evenly spread onto the LB agar or YEM agar plate. Then, the punch out of sterilized clay and control samples (6 mm) were carefully placed on the LB agar plate using a sterile tweezer. Agar plates were incubated overnight at 37 °C incubator, and the study results were recorded by calculating the diameter (mm) of the zones, which was defined by the lack of growth of microorganisms around the respective samples. Results from the experiment are presented as mean zone diameter ± standard deviation ($n \geq 2$).

### 2.3.8. Statistical Analysis

Experiments were conducted in triplicate unless otherwise specified. Data are presented as mean ± standard deviation (S.D.). Statistical comparisons were performed using the two-tailed Student's t-test. Differences were considered statistically significant at $p < 0.05$. Significance levels are indicated as follows: ns (not significant) for $p > 0.05$, * for $p < 0.05$, ** for $p < 0.01$, *** for $p < 0.001$, and **** for $p < 0.0001$.

## 3. Results and Discussion

### 3.1. Synthesis Procedure of Clay Composites for Antimicrobial Application

Figure 1 illustrates the synthesis mechanisms for bentonite and sepiolite clay composites and the testing protocol against microbes. Figure 1 (a) depicts the synthesis of bentonite composites; bentonite clay possesses exchangeable sodium ions within its interlayer spaces. These sodium ions were replaced with silver ions for antibacterial application. Similarly, the antifungal bentonite clay composites were synthesized by incorporating TBH, an antifungal drug. The TBH molecules adhere to the bentonite clay surface by intercalation within the clay's interlayer spaces or surface adsorption[25]. In bentonite clay, small zwitterionic betaine molecules were intercalated into the clay layers, facilitating the accommodation of silver ions and TBH molecules by expanding the interlayer spacing in clay. The electrostatic interaction between the positively charged TBH and the negatively charged clay surface drives the adsorption of TBH onto the bentonite clay[25].

Figure 1(b) presents the synthesis mechanism for sepiolite clay composites. Sepiolite clay composites were synthesized by active adsorption of $Ag^+$ ions and TBH molecules onto the rod-shaped sepiolite clay surface. Like bentonite clay, sepiolite clay also possesses a negatively charged surface, enabling electrostatic interactions with TBH, thus facilitating its adsorption[26]. The synthesized clay composites were then tested against the microbes present in skin wounds. Figure 1(c) represents the schematic of the ZOI study of clay composites against microbes.

The inclusion of the TBH molecule disrupted the clay layers, preventing the formation of a free-standing membrane. All the powder samples were compacted into pellets to be used as scaffolds for skin wound infection. The prototype of clay composites as an antimicrobial patch is shown in supplementary Figure S1 and the flexibility of the Ag-BB-CC (M) membrane is shown in supplementary Figure S12.

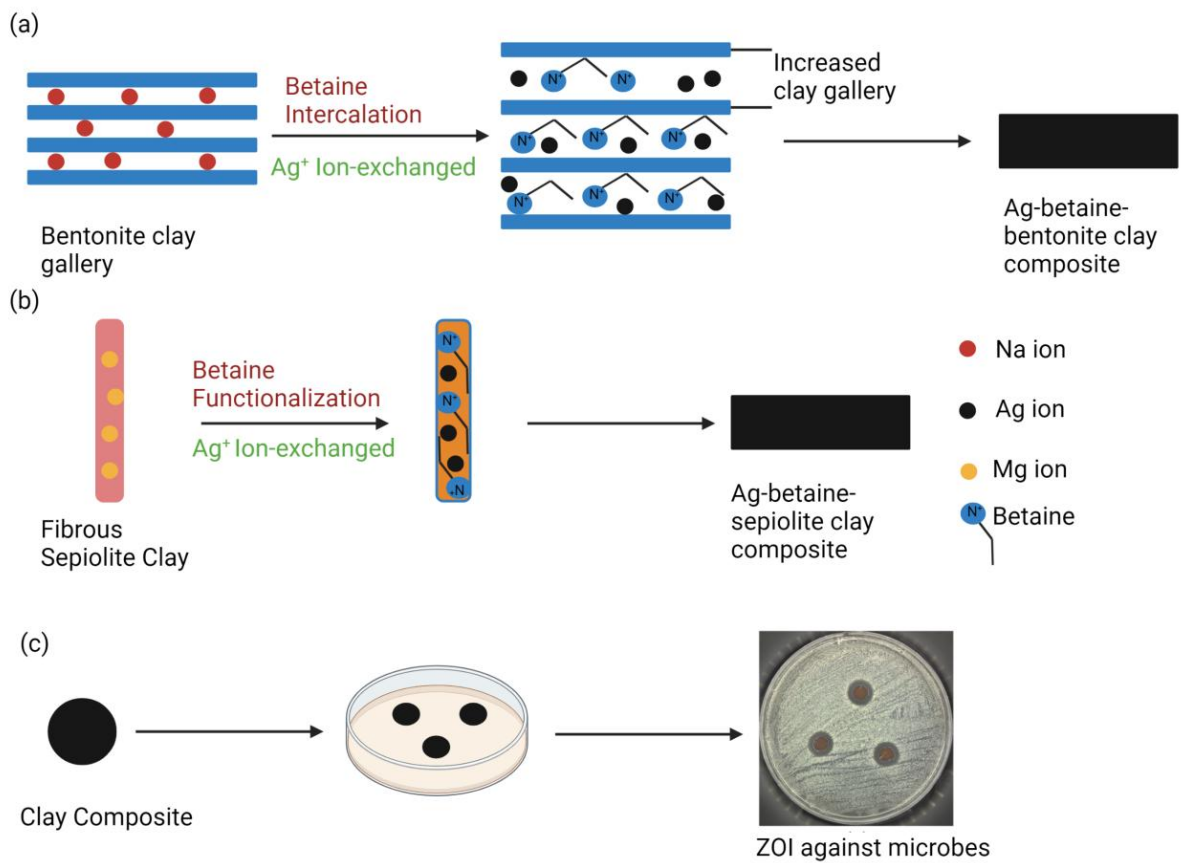

**Figure 1. Schematic representation of functionalized clay composites: (a) Ag-betaine-bentonite clay composite; (b) Ag-TBH-betaine-sepiolite clay composite; (c) Overview of the antimicrobial study conducted on the synthesized clay composites.**

### 3.2. Bentonite Clay Composite

#### 3.2.1. Characterization of Bentonite Clay Composites

Powder X-ray diffraction (XRD) patterns of the samples reveal the intercalation of betaine and silver ions in the bentonite clay interlayers, as shown in Figure 2 (a). The intercalation of the organic species and ion exchange with clay is verified by the change in the $d_{001}$ diffraction peak of the bentonite clay[27]. The $2\theta$ peak for pristine bentonite clay at 7.72° is attributed to the $d_{001}$ diffraction peak, corresponding to a $d$-spacing of 1.15 nm, consistent with the previously published result[28]. The $d_{111}$, $d_{103}$, and $d_{211}$ diffraction peaks of bentonite clay appeared at 20.19°, 26.9°, and 35.3°, respectively[28,29]. For the bentonite-betaine clay composites, BB-CC (M), the $d_{001}$ diffraction peak ($d$ = 1.95 nm) shifted towards a lower angle of 4.54°, exhibiting intercalation of the betaine molecules in the clay galleries with increased $d$-spacing. The increased $d$-spacing resulting from betaine intercalation is influenced by the orientation and packing of betaine molecules within the clay gallery. A $d$-spacing of 1.95 nm suggests the formation of pseudotrimolecular arrangements of betaine in the gallery[30].

However, when silver ions ($Ag^+$) were incorporated into clay, the $d$-spacing in the Ag-Betaine-Bentonite composites (BB-CC (M) decreased. Among the Ag-BB-CC (M) composites, the lowest silver-loaded clay membrane (Ag (1)-BB-CC (M)) showed the highest $d$-spacing of 1.84 nm compared to the highest silver-loaded clay membrane, Ag (4)-BB-CC (M) ($d$-spacing 1.39 nm).

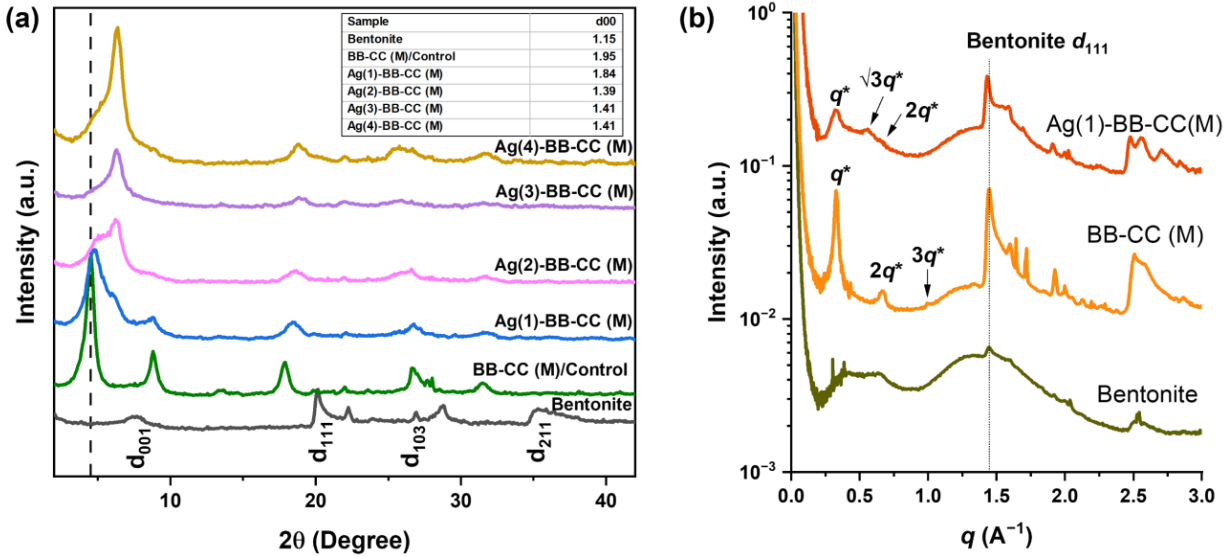

**Figure 2. (a)** X-ray diffraction (XRD) patterns of bentonite clay, BB-CC (M) control sample, and silver-loaded composite membranes (Ag-BB-CC (M)), illustrating structural modifications; **(b)** Small-angle X-ray scattering (SAXS) analysis comparing the nanoscale structural features of bentonite clay, BB-CC (M) control, and Ag (1)-BB-CC (M) composite membranes, revealing morphological differences upon silver incorporation.

The decrease in $d$ spacing could be due to the rearrangement and lower density of the betaine molecules and replacement of the interstitial $Na^+$ ions present in the clay during the ion exchange process with bigger $Ag^+$ ions, which validates the presence of a higher density of silver ions inside the clay gallery[31]. SAXS measurements were performed to further elucidate the nanostructures of bentonite clay nanocomposites (Figure 2 (b)). The bentonite powder showed two broad scattering peaks at low $q$, which indicates the formation of some mesoscale ordered structure. However, the ratio of $q$ values of the two peaks did not correspond to any simple ordered structures (e.g., lamellar, cylinder, body-centered cubic). Some diffraction peaks observed from XRD measurements were present, albeit weak. The $d_{111}$ peak was observed at $q = 1.446$ Å$^{-1}$, which is close to the value obtained from XRD; $2\theta = 20.19°$ corresponds to $q = 4\pi\lambda^{-1}\sin\theta = 1.427$ Å$^{-1}$. The BB-CC (M) sample exhibited well-delineated Bragg peaks at $q^* = 0.328$ Å$^{-1}$, $2q^*$, and $3q^*$, indicating the presence of lamellar structure with $d$-spacing of 1.92 nm. The $d$-spacing determined from SAXS matches quantitatively with that calculated from the XRD data (1.95 nm). When Ag was added to the system (Ag (1) – BB - CC (M)), the SAXS data changed significantly; The Bragg peaks now appear at $q^* = 0.331$ Å$^{-1}$, $\sqrt{3}q^*$, and $2q^*$, indicating the formation of hexagonally packed cylinder structure with $d$-spacing of 1.90 nm and inter-cylinder distance $a = 2.19$ nm. Although the change in $d$-spacing is negligible in SAXS, the structural change from lamellar to cylindrical upon adding Ag was clear.

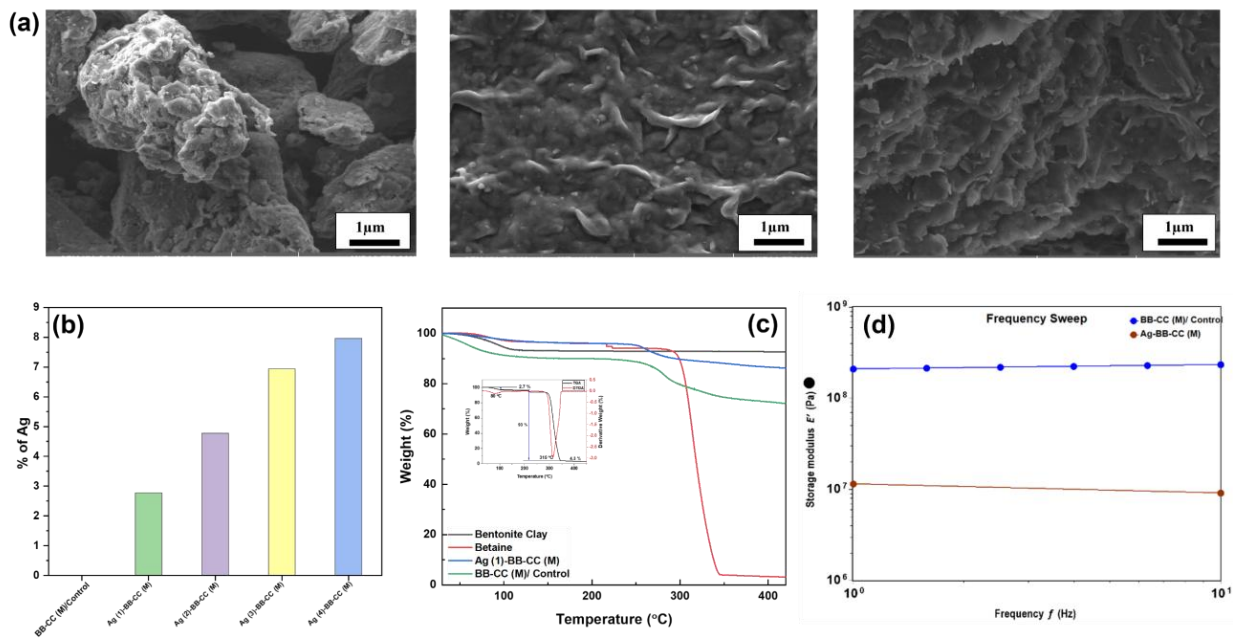

**Figure 3. (a) SEM micrographs of bentonite clay, BB-CC (M)/control, and Ag (1)-BB-CC (M) composite membranes, presented from left to right; (b) ICP-OES analysis quantifying the silver content in Ag-BB-CC (M) composite membranes; (c) Thermogravimetric analysis (TGA) of bentonite clay, BB-CC (M) control, and Ag (1)-BB-CC (M) membranes, evaluating thermal stability; (d) Comparative storage modulus of BB-CC (M)/control and Ag (1)-BB-CC (M), highlighting mechanical performance.**

The structural morphology, textural relationship, and qualitative elemental analyses of the bentonite clay composites, BB-CC (M), and Ag (1)-BB-CC (M) were analyzed by scanning electron microscopy (SEM) and EDS (Figure 3 (a)) (left to right), respectively. The respective particulate and fibrous morphologies of bentonite clay exhibit a transition from aggregated morphology to more arranged and oriented structures, where the platelets come closer, forming a connected network.

ICP-OES analyses were carried out to determine the concentration of $Ag^+$ ions present in the Ag-betaine-bentonite clay composites, and the results are presented in Figure 3 (b). The result showed an increased silver ion concentration from 2.78 % for Ag (1)-BB-CC (M) to 7.96 % for Ag (4)-BB-CC (M), a significant increase that confirmed the successful silver ion in clay composites, which is supported by Energy Dispersive Spectroscopy (EDS) analysis (Figure S3) of pristine bentonite clay, BB-CC (M), and Ag (1)-BB-CC (M). The thermal stability and the degradation of composite membranes at higher temperatures were analyzed using thermogravimetric analysis (Figure 3 (c)). The bentonite clay shows a marked weight loss at ~ 105 °C (9% weight loss), suggesting the dehydration of clay[32]. Dehydration occurs in clay at lower temperatures mainly due to water loss from the clay's external and internal structure. The water absorbed in clay is primarily associated with cations occupying the exchangeable site between clay layers or cation solvation shells[33]. The betaine molecule, a key component of our study, exhibits three distinct decomposition steps (Figure 3 (c) inset). These steps provide a detailed understanding of its thermal behavior. The first step involves a smooth weight loss (2.7%) due to moisture evaporation at 60 - 110 °C. This is followed by a significant weight loss (93%) due to pyrolysis in the 285 - 345 °C range. The final step is a smoother degradation at higher temperatures when the remaining material turns to carbon residue.[34]. It is observed that after dehydration, there is negligible weight loss between 105 and 500 °C for bentonite clay; however, considerable weight loss was observed in clay composite membranes in the range 245-340 °C up to 450 °C. This suggests that the intercalated betaine molecules in the interlayer spaces of bentonite clay have been pyrolyzed and decomposed after the samples were heated to 450 °C. This result agrees with the TGA of the betaine molecule.

Furthermore, the storage modulus of bentonite clay composite membranes was measured using a TA Instruments HR20 rheometer with a DMA attachment (Figures 3 (d) and supplementary Figure S11 in SI). The membranes were cut into 29 mm ×11 mm rectangular pieces. The membrane thickness was 0.05 mm. Frequency sweep test was performed at a strain of 0.01% within the frequency range of 0.1 Hz to 10 Hz at room temperature (25 °C). Minimal deformation was applied to ensure that the material remained within its linear viscoelastic region (LVR), where its properties did not change with deformation[35].

The storage modulus $E'$ indicates the material's ability to store energy elastically and provides information about the stiffness and rigidity of the materials[36]. The BB-CC (M) and Ag-BB-CC (M) membranes exhibit relatively constant storage moduli across the frequency range, which confirms that they behave more elastically, implying a consistent performance of the membranes within the frequency range. The higher storage modulus of BB-CC (M) compared to Ag-BB-CC (M) indicates that BB-CC (M) is more resistant to deformation. We presume the lower storage modulus of Ag-BB-CC (M) is due to modifications in the clay microstructures caused by adding silver ions and a lesser concentration of betaine molecules that dictate the strength and flexibility of the membranes from our earlier investigation based on the betaine-clay viscoelastic properties[22].

ATR-IR analyses (Figure 4) were conducted to explore the IR fingerprints of the functional groups present in the composites and membranes. Figure 4 (a) depicts the ATR-FTIR spectrum of bentonite clay, betaine, and functionalized clay membranes. The broad adsorption peak of bentonite clay is observed at 1000 cm$^{-1}$, corresponding to the Si-O stretching vibration of the tetrahedral sheets in bentonite clay[29]. Other notable bentonite clay peaks are 3625 cm$^{-1}$, 1640 cm$^{-1}$, and 525 cm$^{-1}$. The O-H stretching of the structural hydroxyl group associated with the octahedral layer of the clay correlates to the vibration peak at 3625 cm$^{-1}$[37]. Furthermore, 1640 cm$^{-1}$ and 525

cm$^{-1}$ vibration peaks are assigned to the O-H and Al-O bending of bentonite clay, respectively[32]. The symmetric and asymmetric bending vibrations for COO$^-$ groups are assigned to the new peaks at 1480 cm$^{-1}$ and 1630 cm$^{-1}$ in clay composites, which are absent in bentonite clay. The 1400 cm$^{-1}$ band represents the C-N bond peak of the betaine molecule. This demonstrates the successful intercalation of betaine to the clay gallery[38]. Similarly, the XPS analysis was conducted on pristine bentonite and clay composites to demonstrate the successful functionalization and ion exchange in bentonite clay. The comprehensive spectrum revealed the presence of carbon (C), oxygen (O), silicon (Si), aluminum (Al), sodium (Na), and magnesium (Mg). Notably, N1s peaks emerged in the BB-CC (M) and Ag (1)-BB-CC (M) samples, absent in the pristine bentonite clay (Figure 3 (b)). This signifies the effective functionalization of bentonite clay with betaine molecules. Similarly, Ag peaks in the Ag (1)-BB-CC (M) indicate a successful ion exchange process. For bentonite clay composites, the carbon peaks were analyzed and assigned to specific components at 284.6 eV, 286.5 eV, and 288.6 eV, corresponding to sp$^3$ hybridization of C-C, C-O in the hydroxyl group, and -COOH in the carboxyl group, respectively[39] (Figure S2 and Table S1). The Ag 3d deconvolution in the Ag (1)-BB-CC (M) shows two spin-orbit doublets of Ag 3d$_{5/2}$ and Ag 3d$_{3/2}$ with binding energy values of 367.4 eV and 373.4 eV, respectively. This splitting of 6.0 eV aligns with silver data, suggesting the presence of silver[40]. However, the slight shift in the oxidation state of silver, which has a standard binding energy at around 368.2 eV, is possibly from the electrostatic interaction that may occur between Ag$^+$ and betaine[40]. Additionally, the deconvoluted nitrogen peaks at 402.8 eV and 399.5 eV assigned to CH$_3$-N$^+$ and C=N confirm the interaction of the amine groups present in betaine molecules through the effective functionalization of clay interlayers[41].

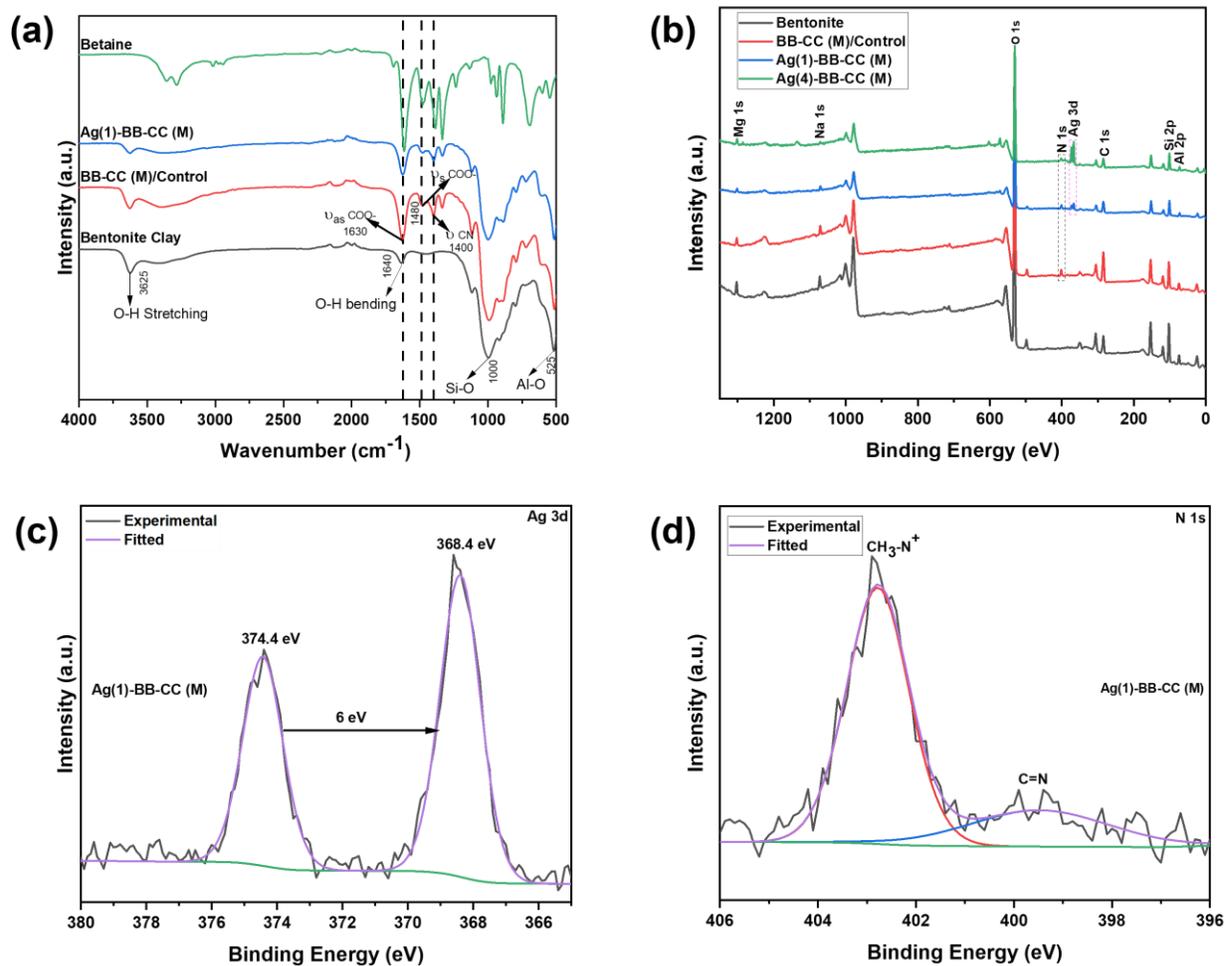

**Figure 4. (a) FTIR spectra of bentonite clay and organofunctionalized clay membranes, (b) Survey spectrum of Bentonite clay, BB-CC (M)/Control, Ag (1)-BB-CC (M), and Ag (4)-BB-CC (M), (c) Ag 3d spectrum of Ag (1)-BB-CC, (d) N1s spectrum of Ag (1)-BB-CC (M).**

The TBH molecules were incorporated into the bentonite clay matrix and betaine to synthesize bentonite-functionalized clay composites with antifungal properties. The corresponding characterization of the synthesized composite is shown in Figure 5, which shows the XPS survey spectrum of TBH-bentonite clay composites at varying TBH concentrations. The presence of respective binding energy (B.E.) peaks for N1s and Cl2p in BB-CC (M) around 400 eV and 198.5 eV in the TBH composites confirm the successful functionalization of the betaine and TBH molecules with the bentonite clay (Figure 5(a)).

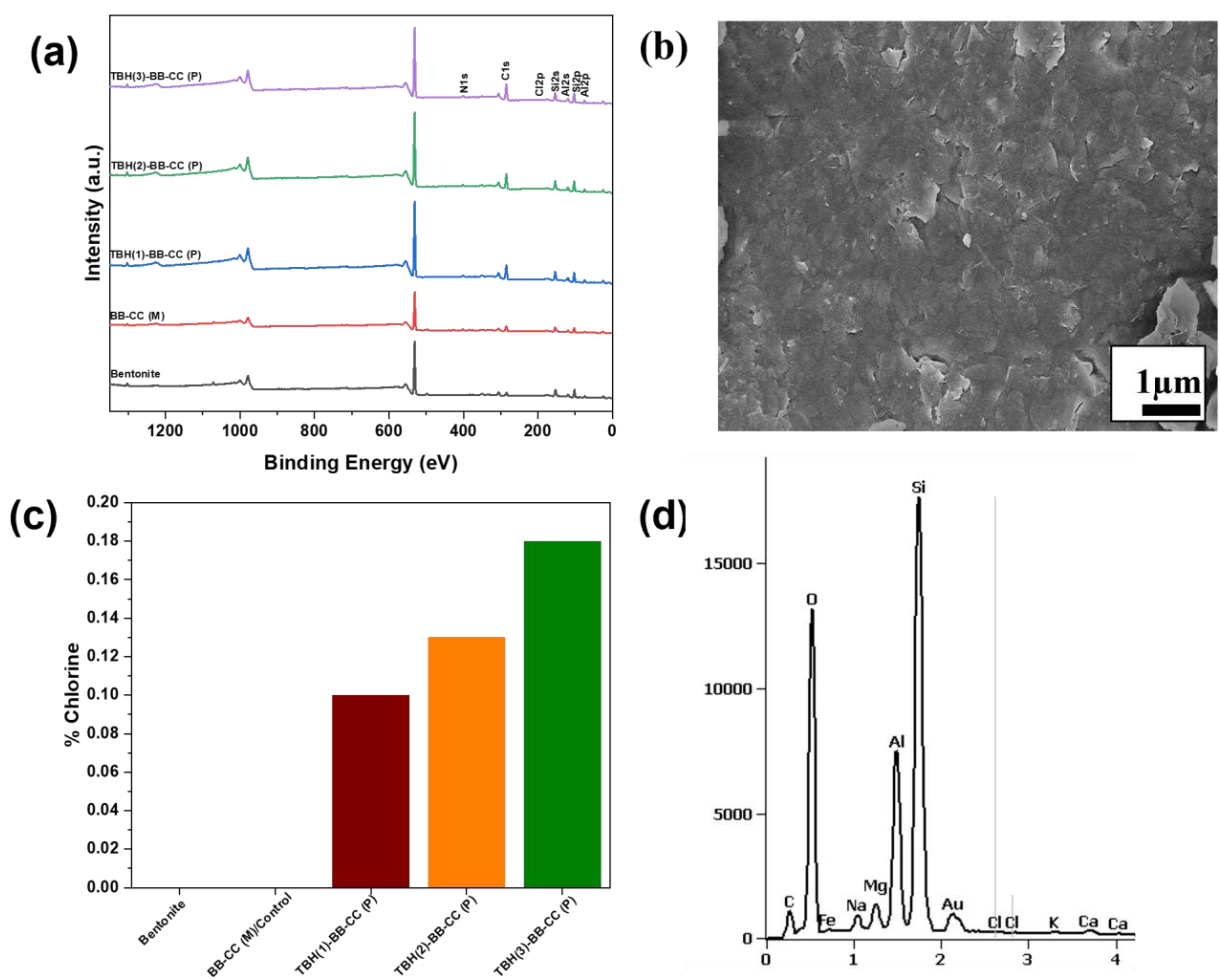

**Figure 5.** (a) XPS survey spectrum of TBH-betaine-bentonite clay composites; (b) SEM image of TBH (1) – BB – CC (P) (c) percentage of chlorine present in TBH-betaine-bentonite-clay composites calculated from XPS survey spectrum; and (d) EDS analysis of TBH (1)-BB-CC (P).

The percentage of chlorine in the TBH composites was calculated from the XPS survey spectrum. From Figure 5 (c), the percentage of chlorine follows the order TBH (3)-BB-CC (P) > TBH (2)-BB-CC (P) > TBH (1)-BB-CC (P), suggesting higher TBH concentration in the same order. This finding aligns with the initial loading of the TBH. Furthermore, the presence of chlorine peaks in EDS analysis confirms TBH functionalization in bentonite clay (Figure 5 (d)). These findings collectively demonstrate the successful functionalization of bentonite clay with TBH molecules synthesized for antifungal application.

A bentonite clay composite membrane with TBH molecule and $Ag^+$ ions was also synthesized to impart antibacterial and antifungal properties. The characterization of the synthesized membrane is presented in Figure 6. Figure 6 (a) represents the FTIR spectra of the TBH, BB-CC (M)/control sample, and Ag-TBH-BB-CC (P) composites. The FTIR spectrum of bentonite clay displays characteristic Si-O peaks at 1000 cm$^{-1}$ and the stretching and bending vibrations of O-H from coordinated water at 3630 cm$^{-1}$ and 1645 cm$^{-1}$, respectively[28]. The FTIR spectrum of TBH exhibits aliphatic C-H stretching[42] at 2975 cm$^{-1}$ and C=C stretching at 1515 cm$^{-1}$. The emergence of new peaks in the BB-CC (M)/ control and Ag-TBH-BB-CC (P) composites can be attributed to betaine modification. These characteristic betaine peaks are observed at 1400 cm$^{-1}$, 1480 cm$^{-1}$, and 1630 cm$^{-1}$, corresponding to C-N stretching and symmetric and asymmetric COO$^-$ stretching, respectively. The appearance of chlorine and Ag peaks in the EDS analysis (Figure 6 (c)) of the Ag-TBH-BB-CC (P) composite confirms the functionalization of Ag and TBH on bentonite clay.

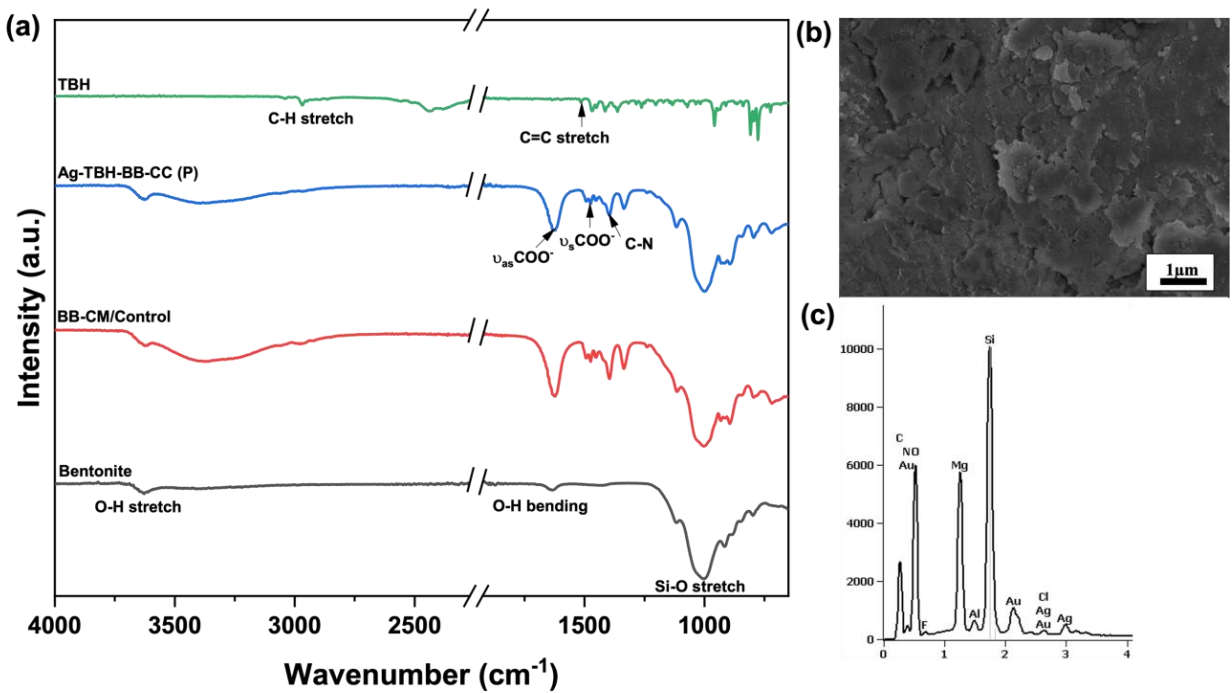

**Figure 6. (a) FTIR analysis of Ag-TBH-BB-CC (P) composite; (b) and (c) SEM image and corresponding EDS analysis of Ag-TBH-BB-CC (P) composite, respectively.**

### 3.2.2. Antimicrobial Activity of Bentonite Clay Composite

The antibacterial activity and efficacy of Ag-BB-CC (M)clay composites were evaluated against *S. aureus* and *E. coli*, two prominent pathogens frequently found in wound infections[43]. Figures 7 (a) and (b) show the ZOI against S. aureus and E. coli, respectively. The size of the inhibition zone signifies the antimicrobial agent's efficacy and is commonly used as a test of resistance against bacteria. For this, the cultures of *S. aureus* and *E. coli* were plated, and bentonite clay composite samples (Ag-BB-CC (M)) with varying amounts of $Ag^+$ were introduced to the agar plate with cultures. After 24 hours of incubation, the Ag-BB-CC (M) composites showed a significant ZOI against *S. aureus* and *E. coli*, which was absent in the control sample (BB-CC (M) composite with no $Ag^+$) (Figure 7 (a)-(b), and Figure S4 and Figure S5). $Ag^+$ impedes bacterial growth by disrupting bacterial membranes. Positively charged $Ag^+$ binds to the negatively charged sulfhydryl group of bacterial membranes through electrostatic interaction[44]. The accumulation of $Ag^+$ leads to protein deactivation and respiratory chain destruction, ultimately hindering bacterial growth and DNA replication[44]. Moreover, the silver ions act as a catalyst to release excessive reactive oxygen species (ROS), inhibiting bacterial growth and proliferation[45]. The evident zones on Ag-BB-CC (M) clay composites show their effectiveness against Gram-positive and Gram-negative bacteria[43,46]. Although these particles can exhibit broad-spectrum antibacterial activity, Gram-positive bacteria remain more susceptible to the action of antibacterial agents compared to Gram-negative bacteria mainly because of the fundamental difference in their structural makeup[47]. Gram-positive bacteria lack an outer protective membrane, which makes them more susceptible to antibacterial agents[47]. In contrast, gram-negative bacterial cell walls possess complex arrangements that make them less susceptible to cidal efficacies than gram-positive bacteria[25]. This may explain the higher inhibition zone diameter in *S. aureus* than in *E. coli*. These results are consistent with the previously published results of silver clay composites[36, 47].

Similarly, the antifungal efficacy of TBH-betaine-bentonite (TBH-BB-CC (P)) clay composites was investigated against *C. albicans*, a yeast commonly found in mucosal and skin surfaces. The control sample (BB-CC (M)) demonstrated no evident ZOI. In contrast, all the TBH-loaded clay composites demonstrated a distinct ZOI against *C. albicans*. Notably, the highest TBH-loaded composite (TBH (3)- BB-CC (P)) exhibited the largest inhibition zone, measuring 18.3 mm (Figure 7 (c) and Supplementary Figure S6). This observation of the direct correlation between TBH loading and antifungal effectiveness is supported by the percentage of chlorine detected in the membrane via XPS survey spectrum analysis (Figure 5 (c)). The antifungal mechanism of TBH comes from its ability to inhibit the fungal squalene epoxidase, which leads to the deficiency in ergosterol[25]. Ergosterol is a crucial component of the fungal cell membrane. The deficiency of ergosterol disrupts the fungal cell membrane. Simultaneously, the accumulation of squalene interferes with the functionality of the fungal cell membrane, ultimately leading to the death of the fungal cells[48].

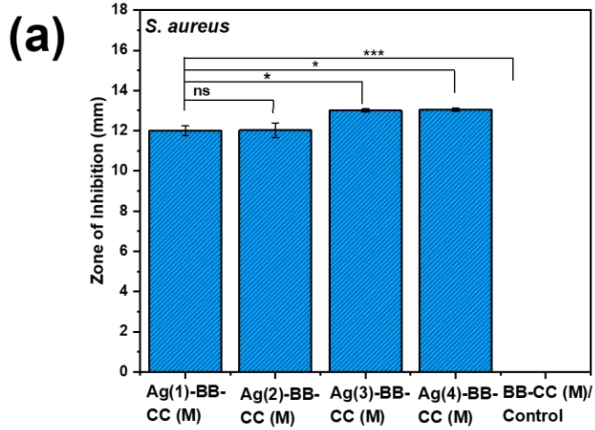
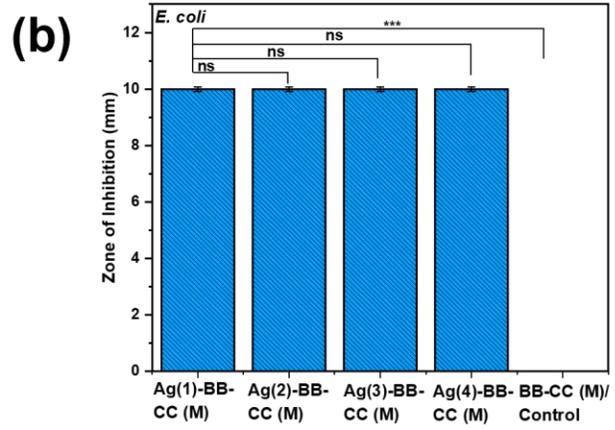
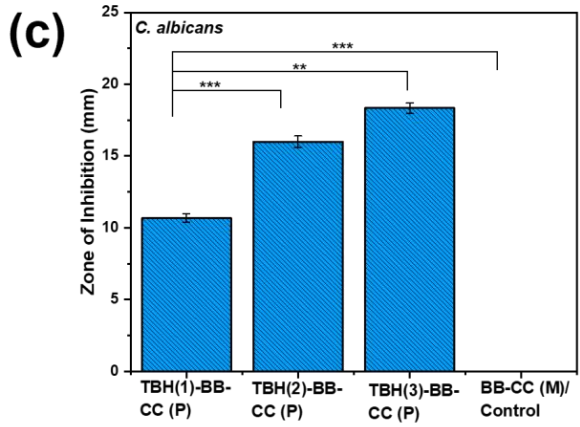
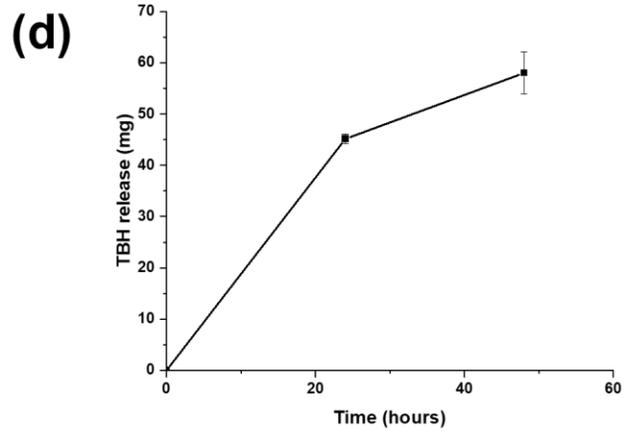
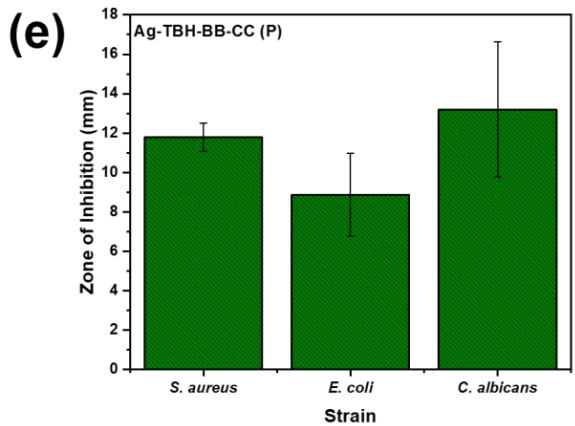
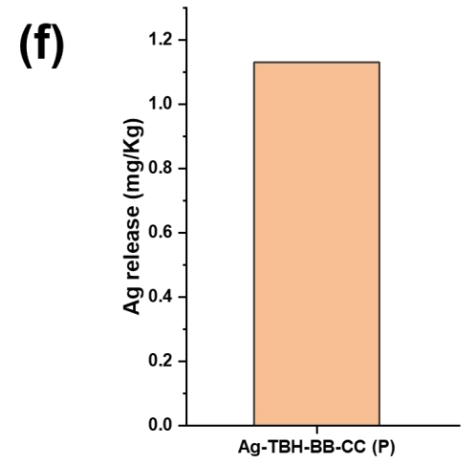

**Figure 7. (a) and (b) Zone of inhibition (ZOI) of Ag-BB-CC (M) against *S. aureus* and *E. coli*, respectively (N = 3). (c) and (d) ZOI of TBH-BB-CC (P) and the TBH release profile from the TBH-BB-CC (P) composite (N = 3). (e) and (f) ZOI of Ag-TBH-BB-CC (P) and the release of Ag$^+$ ions from the membrane after 24 hours.**

Figure 7 (d) represents the TBH release from the TBH (1)- BB-CC (P) sample. The sample was dialyzed in DI water at 37 °C, and the dialysis solution was obtained at 24 and 48 h time points to examine the TBH elution by measuring the absorbance at 283 nm via UV-vis analysis. The release of TBH from the TBH (1)-BB-CC sample gradually reached a plateau by the 48-hour mark, demonstrating controlled release kinetics. This controlled release behavior highlights the potential of the synthesized composites for antimicrobial applications in wound management[49,50].

It is evident from the earlier studies that Ag$^+$ can inhibit the growth of *S. aureus* and *E. coli*, while TBH is effective against *C. albicans*. In this study, the bentonite clay composites developed by infusing Ag$^+$ and TBH into the bentonite clay matrix were tested for their antibacterial and antifungal efficacy on *S. aureus*, *E. coli*, and *C. albicans* and the corresponding ZOI presented in Figure 7 (e) and supplementary Figure S7. The findings demonstrate the Ag-TBH-BB-CC (P) clay composite's effectiveness against bacteria and fungi. Notably, the zone of inhibition for the composite against *S. aureus* is more significant than that against *E. coli*. This could be due to the lack of a protective outer membrane in gram-positive bacteria[47]. Figure 7 (f) shows the silver released over 24 hours.

## 3.3. Sepiolite Clay Composites
### 3.3.1. Characterization of Sepiolite Clay Composites

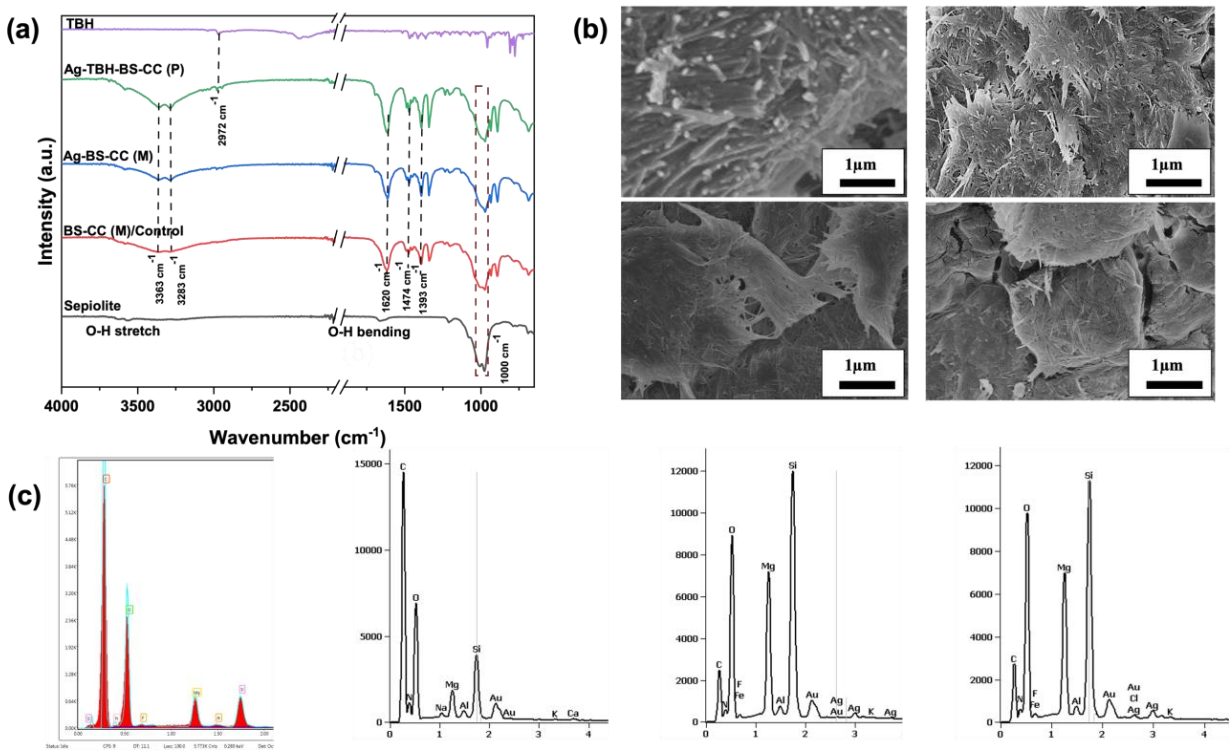

**Figure 8.** (a) FTIR analysis of sepiolite clay composites. (b) SEM images of sepiolite clay and its composites synthesized for antibacterial and antifungal testing: sepiolite clay (top left), BS-CC (M)/control (top right), Ag-BS-CC (M) (bottom left), and Ag-TBH-BS-CC (P) (bottom right). (c) EDS analysis corresponding to the SEM images, presented from left to right: sepiolite clay, BS-CC (M)/control, Ag-BS-CC (M), and Ag-TBH-BS-CC (P).

The FTIR spectra of sepiolite clay and sepiolite clay composites are shown in Figure 8 (a). The Si-O coordination bands at 1000 cm$^{-1}$ and 1213 cm$^{-1}$ are attributed to the Si-O stretching of the Si-O-Si groups in tetrahedral sheets of sepiolite clay[51]. Peaks at 3282-3584 cm$^{-1}$ and 1671 cm$^{-1}$ correspond to the OH stretching and bending of coordinated water, respectively[52]. Similar to the sepiolite clay, new peaks are observed in the BS-CC (M) and Ag-BS-CC (M) clay composites. A peak at 1393 cm$^{-1}$ is assigned to C-N, while peaks at 1483 cm$^{-1}$ and 1620 cm$^{-1}$ are assigned to the symmetric and asymmetric vibrations of the carboxylate (COO$^-$) group[38]. This indicates the successful functionalization of betaine molecules on the sepiolite clay surface. In the Ag-TBH-BS-CC (P) composite, the peaks at 3283-3363 cm$^{-1}$ and 2940-2983 cm$^{-1}$ are from the N-H stretching and aliphatic C-H stretching of the TBH molecules used for functionalization[42].

The high surface area (~ 330 m$^2$/g by N$_2$ BET absorption isotherm), hydrophilicity due to the high density of silanol groups (-SiOH), and high porosity of sepiolite clay make it an excellent material with a high degree of functionalization. The betaine molecules can be functionalized onto the sepiolite surface silanol groups through various mechanisms such as hydrogen bonding, van der Waals forces, and electrostatic interactions[26]. The tethered betaine molecules act as a bridge, binding the fibrous structure of the sepiolite clay[53], which is evident from the SEM images (Figure 8 (b)). EDS analyses of the samples show distinct Ag peaks in the Ag-BS-CC (M) composite and both Ag$^+$ and Cl peaks in the Ag-TBH-BS-CC (P) composite, which are absent in the sepiolite clay and BS-CC (M) control samples. This confirms the successful incorporation of Ag$^+$ into the Ag-BS-CC (M) composite and Ag$^+$ and the TBH molecules into the Ag-TBH-BS-CC (P) composite.

### 3.3.2. Antimicrobial Activity of Sepiolite Clay Composite

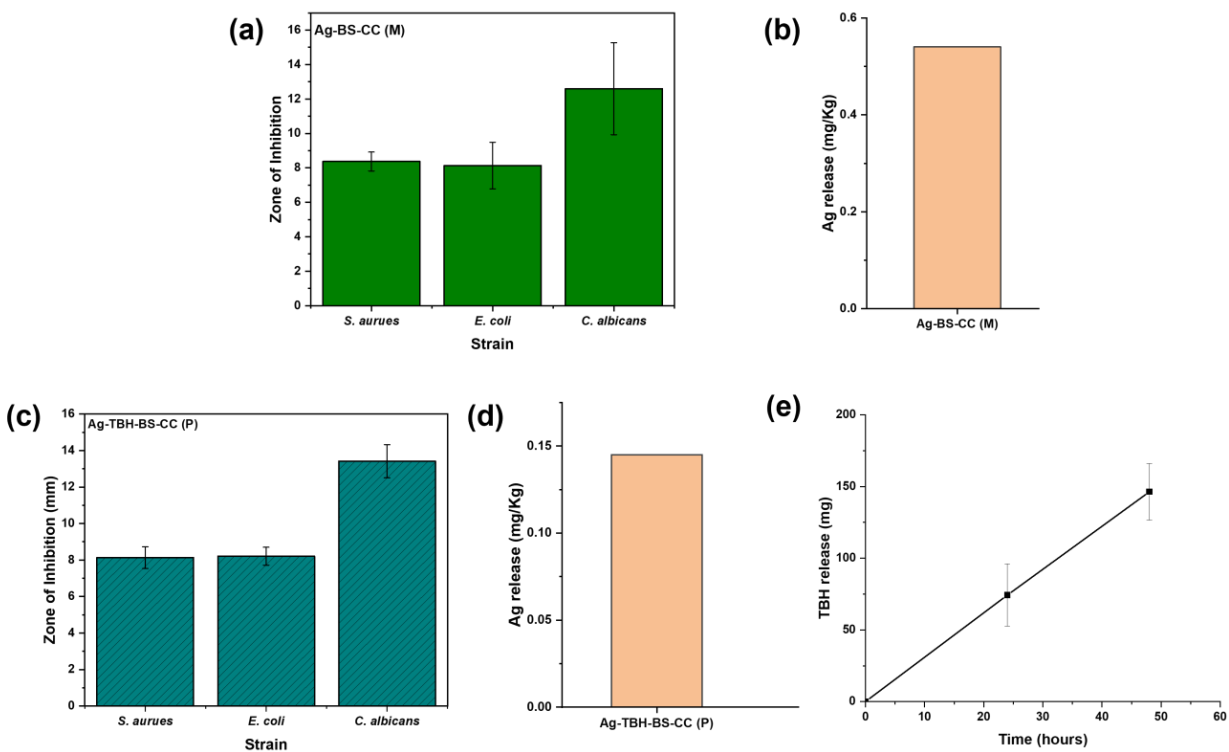

**Figure 9. (a) ZOI of Ag-TBH-BS-CC (M); (b) Ag$^+$ release from Ag-TBH-BS-CC (M) after 24 hours; (c) ZOI of Ag-TBH-BS-CC (P) against *S. aureus*, *E. coli*, and *C. albicans*; (d) & (e) Ag$^+$ and TBH molecule release from Ag-TBH-BS-CC (P) after 24 hours and 48 hours respectively.**

Figure 9 illustrates the antimicrobial efficacy of the synthesized sepiolite clay composites. Figure 9(a) (supplementary Figure S8) shows the clear ZOI of Ag-BS-CC (M) against *S. aureus*, *E. coli*, and *C. albicans*. Figure 9(b) represents the release of silver ions from the composite membranes over 24 hours, with the amount of released silver ions found to be 0.54 mg/kg. These ions contribute to inhibiting microbial growth, as described earlier. Interestingly, unlike the Ag-sepiolite clay composites, the silver bentonite clay composite membranes did not exhibit any inhibition zone against *C. albicans*. This difference is likely due to the higher surface area and porosity of sepiolite clay, which allows more silver ions to attach, enhancing its ability to inhibit the growth of *C. albicans*. Similarly, the ZOI and release profiles of Ag$^+$ and the TBH molecule from Ag-TBH-BS-CC (P) are shown in Figure 9(c–e). The ZOI study (Figure 9(c) and supplementary Figure S9) revealed distinct inhibition zones, suggesting the membrane's effectiveness against *S. aureus*, *E. coli*, and *C. albicans*. Additionally, the release of Ag$^+$ and TBH (Figures 9(d) and 9(e)) from the membrane measured 0.145 mg/kg of silver after 24 hours and 156.22 mg of TBH after 48 hours, indicating a continuous and sustained release of antimicrobial agents from the membrane. The sustained release of antimicrobial agents implies potential applications in wound management, where continuous protection against microbial colonization is crucial for the healing process.

**Conclusion**

This study presents a straightforward and successful method for engineering antimicrobial, functionalized clay composite systems using bentonite and sepiolite clays incorporated with silver ions and terbinafine hydrochloride (TBH). These composite membranes demonstrate excellent efficacy against the pathogens found in skin wounds, such as *S. aureus*, *E. coli*, and *C. albicans*. The research outlines a sequential synthesis process for clay composite membranes for antibacterial, antifungal, and dual-action applications. The study also explored two clays with different morphologies for antimicrobial application, specifically bentonite and sepiolite. Bentonite clay exhibits plate or sheet morphology, while sepiolite clay is characterized by its fibrous or rod-shaped structures. The synthesis techniques were found to be effective for both clay systems. Specifically, clay composite membranes containing silver ions were effective against both gram-positive (*S. aureus*) and gram-negative (*E. coli*) bacteria, and the clay composite membranes with TBH were effective against the fungal strain (*C. albicans*). Membranes containing silver ions and TBH protect against bacterial and fungal pathogens, where silver ions target bacteria and TBH target fungi. On a related note, burn victims have a higher risk and susceptibility to develop such infections. According to a World Health Organization (WHO) report, approximately 11 million people suffer from complications from burn wounds and burn-related infections that cause over 180,000 fatalities globally[54,55]. As such, combating infections has become a health and economic burden worldwide that contributes to increased healthcare expenses and is considered one of the top priorities by the WHO[56]. To circumvent and address issues from pathogenic illnesses, there has been a surge in demand for regenerative medicine in wound management, technological advancements in wound care products, and consideration of alternative approaches that would reduce the overall cost[56]. The global economic impact of the wound care market was $20.8 billion in 2022 and is anticipated to reach $27.2 billion by 2027[56], which propels the development of novel and practical solutions that can exhibit pathogenic resistance and limit microbe penetration and growth within the wound sites in a timely fashion through the advancement and discoveries of materials and drug delivery.

The findings highlight the promising potential of clay-based composite membranes in wound care management. Factors such as biocompatibility, low cost, and low toxicity of clays, combined with their ability to be functionalized with antimicrobial agents, provide an accessible and sustainable approach to treating pathogens in skin wounds. Additionally, the synthesized clay membranes offer sustained release of antimicrobial agents, which are crucial for continuous protection against microbial colonization to reduce the burden of bacterial infections, targeted prevention efforts, and more public health investment. We plan to advance our research to develop antimicrobial treatments for scars, burns, and chronic skin diseases. Future research will optimize and utilize these composites for clinical testing of burns, wounds, and scars using animal models to evaluate their efficacy and effectiveness in real-world applications.

**Author Contributions**

Suvash Ghimire: investigation, formal analysis, data collection, methodology, visualization, writing - original draft, reviewing and editing. Yi Wu: antimicrobial testing, writing: original draft, reviewing, and editing. Manjyot Kaur Chug: antimicrobial testing, writing: original draft, reviewing and editing. Kyungtae Kim: SAXS experiments, resources, methodology, writing: reviewing, editing. Elizabeth J. Brisbois: supervision, resources, project administration, methodology, writing – review and editing. K. Mukhopadhyay: conceptualization, supervision, resources, funding acquisition, project management, analysis, methodology, writing – original draft, writing – review and editing.

**Conflicts of interest**

The authors declare that they have no known competing financial interests.

**Acknowledgments**

ICP-OES analyses were carried out at the Galbraith Laboratories in Knoxville, Tennessee; the ICP-MS analyses were carried out at the Plasma Chemistry Laboratory at the Center for Applied Isotope Studies, University of Georgia. K.M. thanks UCF for a start-up grant; S.G. and K.M. thank the DHS-FEMA grant (EMW-2018-FP00329) for a graduate assistantship for S.G. The authors acknowledge the NSF MRI: XPS: ECCS: 1726636, hosted in the MCF-AMPAC facility at UCF. This work was performed, in part, at the Center for Integrated Nanotechnologies, an Office of Science User Facility operated for the U.S. Department of Energy (DOE) Office of Science. Los Alamos National Laboratory, an affirmative action equal opportunity employer, is managed by Triad National Security, LLC for the U.S. DOE's NNSA, under contract 89233218CNA000001.